%% file: sn-article.tex
\documentclass[pdflatex,sn-mathphys]{sn-jnl}

\jyear{2021}

\theoremstyle{thmstyleone}%

\theoremstyle{thmstyletwo}%

\theoremstyle{thmstylethree}%

\newcommand{\Hb}{\ensuremath{\rm H\beta}}
\newcommand{\Oiii}[1]{[\ion{O}{III}] \ensuremath{#1}}
\newcommand{\OiiiUV}[1]{\ion{O}{III}] \ensuremath{#1}}
\newcommand{\Oiirec}[1]{\ion{O}{II} \ensuremath{#1}}
\newcommand{\Oii}[1]{[\ion{O}{II}] \ensuremath{#1}}
\newcommand{\Cii}[1]{[\ion{C}{II}] \ensuremath{#1}}
\newcommand{\Hii}{\ion{H}{II}}

\newcommand{\secpoint}{\mbox{$''\mskip-7.6mu.\,$}}
\def\arcsecond{$''~$}

\newcommand{\um}{\ensuremath{\mu\mathrm{m}}}
\newcommand\ion[2]{\text{#1\,\textsc{\lowercase{#2}}}}
\newcommand{\chb}{\ensuremath{c(\mathrm{H}\beta)}}
\newcommand{\Rap}{\ensuremath{R_\mathrm{ap}}}
\newcommand{\wabs}{\ensuremath{W_\mathrm{abs}}}
\newcommand{\kms}{\ensuremath{\mathrm{km}/\mathrm{s}}}

\begin{document}

\title[Accurate Abundance in Optical and Far-IR]{Accurate Oxygen Abundance of Interstellar Gas in Mrk 71 from Optical and Infrared Spectra}

\author*[1]{\fnm{Yuguang} \sur{Chen}}\email{yugchen@ucdavis.edu}

\author[1]{\fnm{Tucker} \sur{Jones}}

\author[1,9]{\fnm{Ryan} \sur{Sanders}}

\author[2]{\fnm{Dario} \sur{Fadda}}

\author[2,3]{\fnm{Jessica} \sur{Sutter}}

\author[2,4]{\fnm{Robert} \sur{Minchin}}

\author[1]{\fnm{Erin} \sur{Huntzinger}}

\author[5]{\fnm{Peter} \sur{Senchyna}}

\author[6]{\fnm{Daniel} \sur{Stark}}

\author[7]{\fnm{Justin} \sur{Spilker}}

\author[6]{\fnm{Benjamin} \sur{Weiner}}

\author[8]{\fnm{Guido} \sur{Roberts-Borsani}}

\affil*[1]{\orgdiv{Department of Physics \& Astronomy}, \orgname{University of California, Davis}, \orgaddress{\street{1 Sheilds Avenue}, \city{Davis}, \postcode{95616}, \state{CA}, \country{USA}}}

\affil[2]{\orgdiv{SOFIA Science Center}, \orgname{NASA Ames Research Center}, \orgaddress{\street{M.S. N232-12 Moffett Field}, \city{Mountain View}, \postcode{94035}, \state{CA}, \country{USA}}}

\affil[3]{\orgdiv{Center for Astrophysics and Space Sciences, Department of Physics}, \orgname{University of California, San Diego}, \orgaddress{\street{9500 Gilman Drive}, \city{La Jolla}, \postcode{92093}, \state{CA}, \country{USA}}}

\affil[4]{\orgdiv{Pete V. Domenici Science Operations Center}, \orgname{National Radio Astronomy Observatory}, \orgaddress{\street{P.O. Box O, 1003 Lopezville Road}, \city{Socorro}, \postcode{87801-0387}, \state{NM}, \country{USA}}}

\affil[5]{\orgname{The Observatories of the Carnegie Institution for Science}, \orgaddress{\street{813 Santa Barbara Street}, \city{Pasadena}, \postcode{91101}, \state{CA}, \country{USA}}}

\affil[6]{\orgdiv{Steward Observatory}, \orgname{University of Arizona}, \orgaddress{\street{933 N Cherry Avenue}, \city{Tucson}, \postcode{85721}, \state{AZ}, \country{USA}}}

\affil[7]{\orgdiv{Department of Physics and Astronomy and George P. and Cynthia Woods Mitchell Institute for Fundamental Physics and Astronomy}, \orgname{Texas A\&M University}, \orgaddress{\street{576 University Drive}, \city{College Station}, \postcode{77843-4242}, \state{TX}, \country{USA}}}

\affil[8]{\orgdiv{Department of Physics and Astronomy}, \orgname{University of California, Los Angeles}, \orgaddress{\street{430 Portola Plaza}, \city{Los Angeles}, \postcode{90095}, \state{CA}, \country{USA}}}

\affil[9]{NHFP Hubble Fellow}

\maketitle

\textbf{The heavy element content (``metallicity'') of the Universe is a record of the total star formation history. Gas-phase metallicity in galaxies, as well as its evolution with time, is of particular interest as a tracer of accretion and outflow processes. 
However, metallicities from the widely-used electron temperature ($T_e$) method are typically $\sim 2\times$ lower than the values based on the recombination line method. This ``abundance discrepancy factor'' (ADF) is well known and is commonly ascribed to bias due to temperature fluctuations. 
We present a measurement of oxygen abundance in the nearby (3.4 Mpc) system, Mrk 71, using a combination of optical and far-IR emission lines to measure and correct for temperature fluctuation effects. 
Our far-IR result is inconsistent ( {$> 2 \sigma$ significance}) with the metallicity from recombination lines and instead indicates little to no bias in the standard $T_e$ method, ruling out the long-standing hypothesis that the ADF is explained by temperature fluctuations for this object. 
Our results provide a framework to accurately measure metallicity across cosmic history, including with recent data reaching within the first billion years with JWST and the Atacama Large Millimeter Array (ALMA).
}

Metallicity of the gaseous interstellar medium (ISM) is most easily traced by the abundance of oxygen relative to hydrogen (O/H), measured from emission lines. The most practical method is to determine nebular electron temperature $T_e$ and oxygen abundance from collisionally excited lines (CELs) such as \Oiii{4363,4959,5007}\cite{osterbrock06}, although fluctuations in temperature cause a systematic underestimate of the true abundance\cite{peimbert67}. 
Measurements from recombination lines (RLs) in contrast are nearly insensitive to $T_e$ effects, and the resulting O/H abundances are typically $\sim$0.2 dex (or $\sim$1.6$\times$) higher than those from CELs\cite{peimbert93, blanc15, esteban09, esteban14}. 
Temperature fluctuations offer a natural explanation for this discrepancy, but would imply a relatively large magnitude of fluctuations, and previous attempts to reconcile the two methods have been inconclusive or contradictory \cite{bresolin09, croxall13, stasinska13}. 
Here, we combine optical RL and CEL measurements with far-infrared (far-IR) CEL lines which have the advantage of being comparably insensitive to $T_e$ as the RL emission \cite{osterbrock06}. This allows us to directly measure the temperature fluctuations and correct for any bias in the standard optical CEL result. The aim of our methodology is to investigate the cause of the abundance discrepancy, and to establish an accurate O/H metallicity scale. 

We selected the target Markarian~71 (Mrk~71; also called NGC~2366-I, or mislabeled as NGC~2363 \cite{corwin04}) for a case study based on its physical properties. 
Mrk~71 is one of the most nearby (distance $= 3.4~\mathrm{Mpc}$) analogues to high-$z$ star-forming galaxies, with high electron temperature ($T_e \sim 15,000~\mathrm{K}$), high ionisation parameter ($\log U \sim -2.2$), and compact morphology, which are consistent with young starbursts (stellar age of the primary western component is $\sim 1~\mathrm{Myr}$) \cite{micheva17}. Such properties are extreme among nearby galaxies, but common in star-forming galaxies at $z \gtrsim 2$ \cite{steidel14, tang19, sanders20, endsley21} and similar to recent JWST-based measurements at $z\gtrsim8$\cite{cordova22, katz22, curti22}.
Moreover, low dust attenuation in Mrk 71 makes it especially suitable for comparing optical and infrared (IR) fluxes. 

\begin{figure}[htbp]
\centering
\includegraphics[width=9cm]{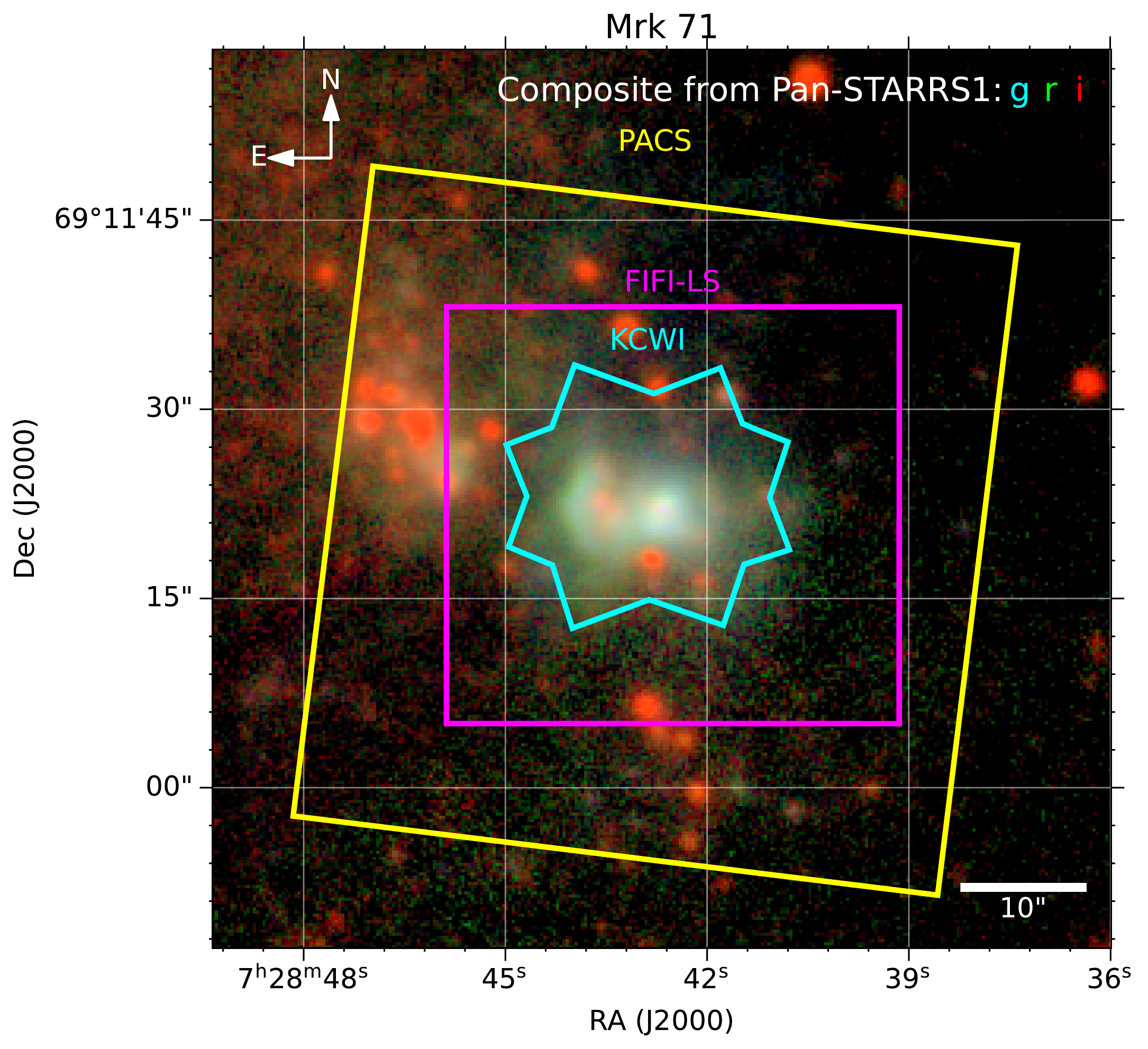}
\caption{ Overview of our spectroscopic datasets on Mrk~71. Solid lines mark the IFU field-of-view boundaries for Herschel/PACS (yellow), SOFIA/FIFI-LS (magenta), and Keck/KCWI (cyan). The background image is a colour composite from Pan-STARRS1 {\it gri} images \cite{chambers16, flewelling20}. The brightest nebular emission corresponds to the diffuse blue emission within the KCWI field. } 
\label{fig:fov}
\end{figure}

We conducted multi-wavelength observations of Mrk~71 using the Keck Cosmic Web Imager (KCWI)\cite{morrissey18} at W.~M. Keck Observatory and the Far Infrared Field-Imaging Line Spectrometer (FIFI-LS)\cite{fischer18} onboard the Stratospheric Observatory for Infrared Astronomy (SOFIA). We additionally use archival observations from the Photodetector Array Camera and Spectrometer (PACS)\cite{poglitsch10} aboard the Herschel Space Observatory. The KCWI observations reveal a wealth of optical emission lines from $\simeq 3500$--5500~\AA, including \Oiii~4363, 4959, and all hydrogen Balmer transitions except H$\alpha$. 
FIFI-LS and PACS data cover the far-IR \Oiii~52~\um\ and 88~\um\ emission lines, respectively. 
All three instruments are integral field spectrographs (IFSs), which provide both spatial and spectroscopic information, allowing us to accurately match the point spread function (PSF) and extraction aperture. 
The fields of view (FoV) of the observations used in this work are presented in Fig.~\ref{fig:fov}. A full description of the observations and data reduction is included in the Methods. 

\begin{figure}[htbp]
\centering
\includegraphics[width=10cm]{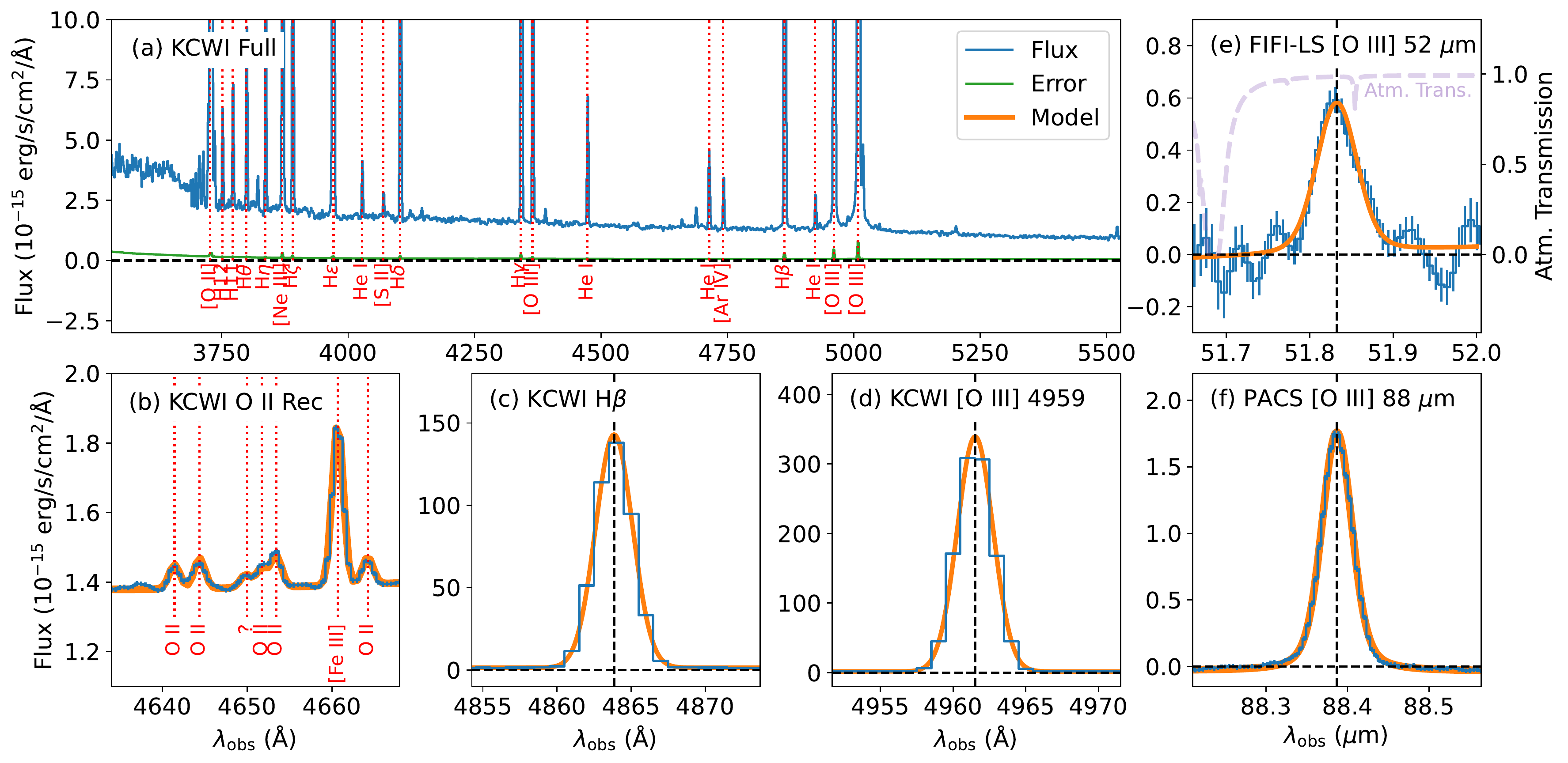}
\caption{  A summary of spectra we obtained for Mrk~71. (a)-(d) Overviews of the KCWI spectra, with (b) extracted from the deep BM exposures and the rest from the shallow BL snapshots. (e) FIFI-LS spectra of \Oiii~52~\um. The atmospheric transmission is also plotted as a function of wavelength; the emission line falls in a region of high transmission. (f) PACS spectra of \Oiii~88~\um. The best-fit emission line models are plotted in orange.   } 
\label{fig:all_spec}
\end{figure}

We extracted PSF- and aperture-matched spectra to measure the flux ratios of optical and far-IR emission lines (Fig. \ref{fig:all_spec}). The line fluxes were measured with 1-D Voigt profile fits to the spectra (Methods). The observed fluxes must be corrected for dust attenuation, which we calculated using the hydrogen Balmer lines from H$\beta$ to H$\kappa$ (H12) (Methods). Assuming a Milky Way extinction curve with $R_V = 3.1$\cite{cardelli89}, the average correction factor within the extraction aperture was measured as \chb~$=0.09\pm 0.04$ (or $A_{\mathrm{H}\beta} = 0.2 \pm 0.1$). Due to the relatively low extinction level, altering the assumed extinction curve does not significantly affect the result (Methods).

\begin{figure}[htbp]
\centering
\includegraphics[width=8cm]{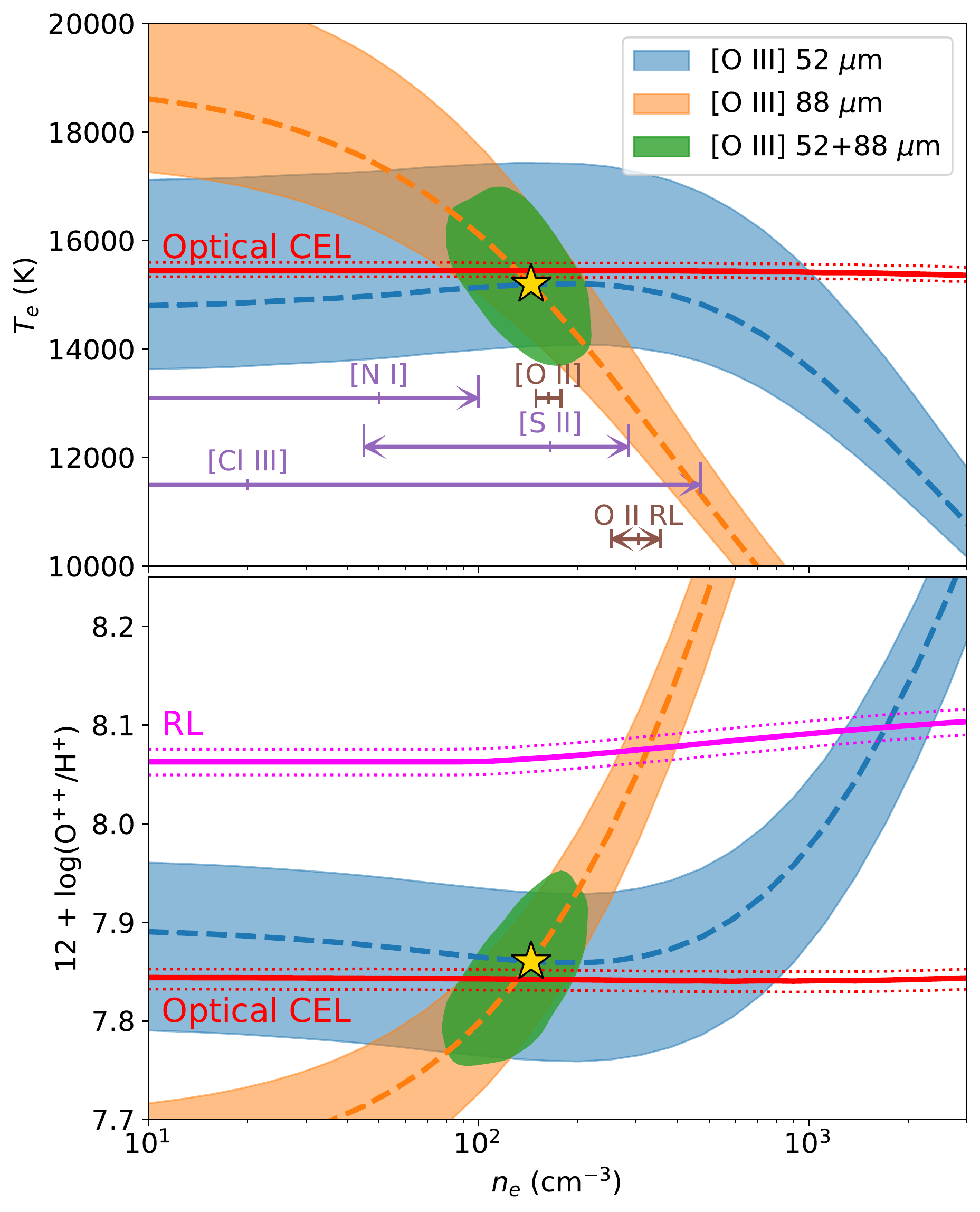}
\caption{  A summary of $T_e$ (top) and O$^{++}$/H$^+$ abundances (bottom) of Mrk~71 measured from different direct methods. The shaded regions and dotted lines indicate 1-$\sigma$ uncertainties. Metallicity derived from the far-IR \Oiii~52~\um\ (blue) and \Oiii~88~\um\ (orange) is sensitive to electron density $n_e$, such that these two lines together provide a joint constraint (green) of $n_e$, $T_e$, and metallicity. The far-IR metallicity is consistent with that measured from optical CEL (red), and $\gtrsim 3\sigma$ discrepant from the RL-based value (magenta). This is contradictory to the temperature fluctuation hypothesis, which predicts far-IR and RL metallicities should be nearly identical. 
We also show $n_e$ measurements from \Oii\ and \Oiirec\ RL measured from our KCWI spectra (brown) along with [\ion{N}{I}], [\ion{S}{II}], and [\ion{Cl}{III}] reported previously \cite{esteban09} (purple). }
\label{fig:oh_metallicity}
\end{figure}

A summary of the physical properties derived from the reddening-corrected line fluxes is provided in Table~\ref{tab:properties} and Fig.~\ref{fig:oh_metallicity}. 
We determined $T_e$ and the CEL-based O$^{++}$/H$^{+}$ metallicity following standard methods using the flux ratios of \Oiii~4959, \Oiii~4363, and \Hb\ (see Methods; \Oiii~5007 is partially saturated in our KCWI data and is not used). 
Far-IR-based metallicities from \Oiii~52~\um\ and 88~\um\ emission lines were calculated similarly, but with $T_e$ determined by the ratios between the far-IR lines and \Oiii~4959, and metallicities measured from the far-IR to H$\beta$ ratios. These far-IR results are less sensitive to $T_e$ \cite{osterbrock06}. 
RL-based O$^{++}$/H$^{+}$ metallicity was measured from the total flux of the \Oiirec~4649 V1 multiplet and H$\beta$. 
We note that 1-$\sigma$ errors herein include both the statistical uncertainty (i.e., noise) and the flux calibration uncertainty across different instruments. 

We find an O$^{++}$ abundance discrepancy of 0.23 dex from the KCWI spectra of Mrk 71 (i.e., the RL abundance is 1.7$\times$ larger than from optical CELs, which is typical of \Hii\ regions). 
If this discrepancy is caused by temperature fluctuations, then the $T_e$ measured from \Oiii~52,88~\um/\Oiii~4959 should be lower than from \Oiii~4363. Additionally, the far-IR CEL metallicity from \Oiii~52~\um, 88~\um\ should be nearly identical to the RL metallicity due to their similar insensitivity to $T_e$. Instead, we find that $T_e$ and O$^{++}$ abundance from the far-IR lines are consistent with the optical CELs, with far-IR metallicity smaller than the RL value at $>2\sigma$ significance (Fig. \ref{fig:oh_metallicity}). 
This result represents an upper bound on the temperature fluctuations present in the ionized (O$^{++}$) gas.

The magnitude of fluctuation in $T_e$ is usually quantified by the dimensionless variance $t^2$\cite{peimbert67}:
\begin{equation}
t^2 = \frac{\int n(\mathrm{O}^{++}) n_e (T_e - T_0)^2 \, dS}{T_0^2 \int n(\mathrm{O}^{++}) n_e \, dS},
\end{equation}
where $n(\mathrm{O}^{++})$ and $n_e$ are the densities of O$^{++}$ ions and electrons, and $T_0$ is the mean electron temperature that can be measured alongside $t^2$. 
Our RL and optical CEL results require a value $t^2 \approx 0.10$ to reconcile the abundances (Table~\ref{tab:properties}), assuming a Gaussian $T_e$ distribution. The comparison of far-IR-to-optical CELs instead gives $t^2 = 0.008\pm0.043$, consistent with zero (no temperature fluctuations), and inconsistent at the  {$> 2\sigma$} level with the value implied by the RL lines. 
We emphasise that this is a {\it direct measurement of $t^2$} from the relevant O$^{++}$ ion emission (as opposed to indirect inference from RLs, or from the Balmer jump temperature\cite{peimbert17}).

We now turn to the question of the true gas-phase metallicity scale. 
The combined optical and IR spectroscopy allows us to make a direct measurement of the O$^{++}$ ion abundance, including a correction for the temperature fluctuation effects, thereby eliminating a potentially large systematic bias in the $T_e$ method. We instead derive a minimal correction: IR measurements imply that the abundance based on the standard \Oiii{4363} diagnostic is underestimated by only $0.02^{+0.07}_{-0.11}$ dex, with a standard deviation in the temperature distribution of $< 3000$ K. 
This result implies that the abundance discrepancy (of 0.23 dex in this case) is not caused by temperature fluctuations, as has been widely suggested, but rather that CEL metallicities are reasonably accurate despite their temperature sensitivity. 
Alternatively, the discrepancy might be explained by fluctuations of density (indicated by various $n_e$ measurements from other ionic species \cite{esteban09}; Fig. \ref{fig:oh_metallicity}) in addition to temperature.  {In this case, a component of gas with $n_e \gtrsim 10^3$~cm$^{-3}$ (i.e., higher than the critical density of \Oiii~52~\um) could cause IR CEL measurements to underestimate the true abundance, while not significantly affecting the optical metallicities. However, values of $n_e$ measured from \Oii\ and other optical features \cite{esteban09} do not support the presence of such a dense component. 
Additionally, in the presence of strong density and temperature fluctuations, we would also expect a significant discrepancy between optical and UV-based CEL measurements of $T_e$. A recent study of the CLASSY sample \cite{mingozzi22} reports that the optical and UV-based $T_e$, on average, differ by only $\sim$1000~K, which is not sufficient for the temperature-fluctuation scenario to explain the observed ADF.} 
Another explanation for the discrepancy is that RL metallicities may be biased high by ``inclusions'' of metal-rich gas occupying relatively small volumes \cite{liu00, stasinska07,tsamis2003}. 
Such a chemical inhomogeneity causes only a small bias in the CEL metallicities, as required by our $t^2$-corrected results.

We caution that our measurements are strictly for the gas-phase metallicity,
 {whereas a fraction of the total oxygen abundance is depleted into solid dust grains. Assuming that the true total oxygen abundance is $\sim$0.1 dex higher than the gas-phase value \cite{jenkins09}, our finding is in agreement with recent comparisons of stellar and nebular abundances \cite{bresolin16, bresolin22}.  Specifically, individual young stars in galaxies such as NGC~300 and NGC~2403 are consistent with nebular $T_e$-based metallicities (accounting for the $\sim$0.1~dex depletion). While the actual amount of depletion is under debate, these stellar measurements further support our conclusion that the RL metallicity might be biased high while the $T_e$ method is closer to the accurate value. 
}

Our finding of consistent CEL-based metallicities from both optical and IR lines is encouraging for chemical evolution studies reaching into the reionization epoch. While $T_e$ measurements using the auroral \Oiii{4363} lines have been demonstrated at $z>7$ with JWST \cite{cordova22, katz22, curti22}, they will be increasingly difficult at higher redshifts. Meanwhile, the UV \OiiiUV{1661,1666} doublet and IR \Oiii{52~\um, 88~\um} lines are promising targets for JWST and ALMA\cite{jones20, bakx2022}, respectively, at $z\sim6-12$ and beyond. Current Cycle 1 JWST GO programmes are already targeting galaxies for this purpose. A large temperature fluctuation effect would result in systematically higher (lower) metallicities derived from IR (UV) lines compared to optical \Oiii{4363}, complicating interpretations of mixed-wavelength datasets \cite{sanders21}. Instead, our measurements suggest that these various diagnostics can be compared with negligible bias.
Verifying our result with a larger sample -- and with UV emission lines -- is nonetheless an important next step to ensure reliable metallicities 
across cosmic history.

 {\textit{Correspondence and requests for materials should be addressed to Y. Chen.}}

\noindent\textbf{Acknowledgements}

Based on data obtained at the W. M. Keck Observatory, which is operated as a scientific partnership among the California Institute of Technology, the University of California, and the National Aeronautics and Space Administration. The Observatory was made possible by the generous financial support of the W. M. Keck Foundation. The authors wish to recognize and acknowledge the very significant cultural role and reverence that the summit of Maunakea has always had within the indigenous Hawaiian community.  We are most fortunate to have the opportunity to conduct observations from this mountain. Results in this paper are based on observations made with the NASA/DLR Stratospheric Observatory for Infrared Astronomy (SOFIA). SOFIA is jointly operated by the Universities Space Research Association, Inc. (USRA), under NASA contract NAS2-97001, and the Deutsches SOFIA Institut (DSI) under DLR contract 50-OK-0901 to the University of Stuttgart. Herschel is an ESA space observatory with science instruments provided by European-led Principal Investigator consortia and with important participation from NASA. The authors thank the W. M. Keck Observatory staff, the SOFIA observatory staff, and the Herschel Data Archive for making this study possible. 
Financial support for this work was provided by NASA through award \#08\_0071 issued by USRA.
R.S. acknowledges the support by NASA through the NASA Hubble Fellowship grant \#HST-HF2-51469.001-A awarded by the Space Telescope Science Institute, which is operated by the Association of Universities for Research in Astronomy, Incorporated, under NASA contract NAS5-26555. J.Sutter gratefully acknowledges funding from STScI grant JWST-GO-02107.006.A. R.M. acknowledges support of the NRAO. The National Radio Astronomy Observatory is a facility of the National Science Foundation operated under cooperative agreement by Associated Universities, Inc.

\noindent\textbf{Author Contributions}

Y.C. led the overall data reduction, analysis, and interpretation of the project. T.J. and R.S. conceived the project. Y.C. conducted the KCWI data reduction and analysis. D.F. conducted the FIFI-LS data reduction. D.F., J.Sutter and R.M. contributed to the improvement of the PACS data. All authors contributed to the planning of the observations, the overall interpretation of the results, various aspects of analysis, and the preparation of the manuscript.  

\noindent\textbf{Data availability} 

The raw KCWI data in this work are available on the Keck Observatory Archive: \url{https://www2.keck.hawaii.edu/koa/public/koa.php}. The reduced FIFI-LS data are available at the SOFIA data archive: \url{https://irsa.ipac.caltech.edu/Missions/sofia.html}. The PACS data before transient correction are available at the Herschel data archive: \url{http://archives.esac.esa.int/hsa/whsa/}. Additional data can be provided upon reasonable request.

\noindent\textbf{Code availability}

This study uses publicly available software/packages, including: \textsc{PyNeb} \cite{luridiana15}, \textsc{Astropy} \cite{astropy22}, \textsc{emcee} \cite{emcee13}, SOSPEX (\url{https://github.com/darioflute/sospex}). For the KCWI data, the post-DRP background subtraction and stacking code is maintained at: \url{https://github.com/yuguangchen1/kcwi}. Additional analysis code can be provided upon reasonable request. 

\noindent\textbf{Competing Interest Statement}

The authors declare no competing interest.

\newpage

\noindent\textbf{Figures and Tables}

\begin{table}[h]
\begin{center}
\begin{minipage}{174pt}
\caption{Properties of Mrk~71}\label{tab:properties}%
\begin{tabular}{@{}lc@{}}
\toprule
Physical Properties & Values  \\
\midrule
$\chb$ & $0.09 \pm 0.04$\\
\wabs & $0.6 \pm 0.7~\mathrm{\AA}$\\
$n_e$ (far-IR) & $140_{-70}^{+50}~\mathrm{cm}^{-3}$ \\
$n_e$ (RL) & $310 \pm 50~\mathrm{cm}^{-3}$\\
$n_e$ (\Oii)\footnotemark[1] & $160 \pm 10~\mathrm{cm}^{-3}$\\
$T_e$ (far-IR) & $15,200_{-1,000}^{+1,900}~\mathrm{K}$ \\
$T_e$ (CEL)\footnotemark[2]  &   $15,400 \pm 100~\mathrm{K}$  \\
$12 + \log(\mathrm{O}^{++} / \mathrm{H}^+)$ (far-IR) & $7.86_{-0.11}^{+0.07}$\\
$12 + \log(\mathrm{O}^{++} / \mathrm{H}^+)$ (CEL)\footnotemark[2] & $7.842_{-0.011}^{+0.009}$\\
$12 + \log(\mathrm{O}^{++} / \mathrm{H}^+)$ (RL)\footnotemark[2] & $8.07 \pm 0.01$\\
$12 + \log(\mathrm{O} / \mathrm{H})$ (far-IR)\footnotemark[3] & $7.89_{-0.10}^{+0.06}$ \\
$t^2$ (between far-IR and CEL) & $0.008 \pm 0.043$ \\
$t^2$ (between CEL and RL) & $0.097_{-0.009}^{+0.008}$ \\
\botrule
\end{tabular}
\footnotetext[1]{Measured from the \Oii~3726/3729 ratio, assuming $T_e(\Oii) = 0.71 \times T_e(\Oiii~\mathrm{CEL}) + 3,050~\mathrm{K}$ \cite{esteban09}.}
\footnotetext[2]{Assuming $n_e$ from far-IR.}
\footnotetext[3]{Estimated as $\mathrm{\frac{O}{H} = \frac{O^{+}~+~O^{++}}{H^{+}}}$, where the O$^+$ abundance is measured from reddening-corrected \Oii~3726, 3729 fluxes. }
\end{minipage}
\end{center}
\end{table}

\newpage

\section*{Methods}
\textbf{Keck/KCWI Observations and Data Reduction}

\renewcommand{\thetable}{S\arabic{table}}  
\renewcommand{\thefigure}{S\arabic{figure}}
\setcounter{figure}{0}
\setcounter{table}{0}

The Keck/KCWI data of Mrk~71 were obtained on 2022 March 3 under clear conditions with a seeing of $\lesssim 1''$. All exposures were obtained using the ``Medium'' image slicer, which offers a FoV of approximately 16\secpoint5 $\times$ 20\secpoint4, and a spatial sampling of 0\secpoint64 $\times$ 0\secpoint29. Two snapshots, each with 4 seconds integration, were obtained with the BL grating centred at 4500~\AA, which has a spectral range of $\simeq 3500$--5500~\AA\ and resolution $R \sim 1500$. 
These exposures were taken to capture the bright emission lines, in particular H$\beta$ and \Oiii~4959, while avoiding saturation. A sequence of long exposures with integration time of $600~\mathrm{s} \times 7$ was taken with the BM grating, centred at 4400~\AA, with a spectral range of $\simeq 3960$--4840~\AA\ and $R\sim 3000$. These deep exposures with higher spectral resolution were taken to detect and resolve the faint \Oiirec\ recombination lines. 

The KCWI observations were reduced with the Python version of the Data Reduction Pipeline (DRP)\footnote{\url{https://github.com/Keck-DataReductionPipelines/KCWI_DRP.git}}. For each exposure, the DRP conducts customisable procedures including cosmic ray removal, flat-fielding, wavelength calibration, sky subtraction, atmospheric dispersion correction, and flux calibration. The DRP generates a rectified data cube. However, the default resampling scheme in the DRP uses cubic interpolation that introduces a ``rippling'' feature near bright emission lines. We therefore modified the DRP to use linear interpolation to preserve the spectral profiles and, most importantly, the fluxes of emission lines. For sky subtraction, b-spline sky models were built from the exposures on a blank field and subtracted from the on-target exposures. However, due to sky variability, over- or under-subtractions may occur in this method. We manually subtract the continuum emission of the faintest spaxels from the cube, leading to slight oversubtraction. However, the continuum flux is not used in this work, except during the reddening correction (see below for a detailed discussion). For flux calibration, we observed two white dwarfs (WD), G~47-18 and GD~190, as photometric standards at the beginning and end of the night. The inverse sensitivity curves derived from multiple exposures were averaged and applied to all data cubes. We used a post-DRP process described in \cite{chen21} to combine multiple exposures into final data cubes. For each instrument configuration, the process corrects the astrometry of each exposure with cross-correlation, resamples individual cubes onto a common 0\secpoint3 $\times$ 0\secpoint3 grid using the ``drizzle'' algorithm provided by \textsc{Montage}\footnote{\url{http://montage.ipac.caltech.edu/}}, and averages the frames weighted by exposure times.

\noindent\textbf{SOFIA/FIFI-LS Observations}

SOFIA \cite{temi2018} observations were taken with the Field-Imaging Far-Infrared Line Spectrometer \cite{colditz18, fischer18} (FIFI-LS) during Cycle 8 as part of program 08\_0071. 
FIFI-LS has two channels, each with a 5$\times$5 array of spaxels. The FoV is 30\arcsecond $\times$ 30\arcsecond in the blue channel and 60\arcsecond $\times$ 60\arcsecond in the red channel. We simultaneously targeted \Oiii{52~\um} and \Cii{158~\um} in the blue and red channels, with spectral resolving powers of $R=950$ and 1150 respectively. 

Data were obtained on a single flight leg of approximately 2 hours duration on 2021 April 14 at altitudes of 42,000--43,000 feet. 
We used a standard symmetric chop between the target and blank sky regions, with a 225$^{\circ}$ position angle and 60\arcsecond distance. 
Subpixel dithering was performed with a 9-point symmetric grid pattern of $\pm$4\arcsecond offsets in order to better sample the point spread function. We thus obtained full-depth coverage of \Oiii{52~\um} emission within a 26\arcsecond $\times$ 26\arcsecond region. The dither pattern was repeated multiple times, reaching a total on-source integration time of 2950 seconds.

Our analysis is based on the Level-4 data product from the default reduction process\footnote{\url{https://github.com/SOFIA-USRA/sofia_redux}}. To obtain accurate line fluxes, the atmospheric transmission was recomputed in \textsc{SOSPEX} using only the transmission at the centre of the \Oiii~52~\um\ emission.

\noindent\textbf{Herschel/PACS Archival Data and Reprocessing}

The Herschel/PACS data were obtained from the Herschel Science Archive (HSA). The observation (Observation ID: 1342220604; PI: E. Sturm) was conducted in the B2B band with a spectral coverage of 87.793--89.004~\um. The observation has a total exposure time of 795 seconds and was carried out in the so-called ``unchopped'' mode. Compared to the commonly used ``chop-nod'' mode, these observations were made without a nearby empty reference field. The calibration of unchopped data suffers from detector response variations (``detector transients''), which results in a systematic bias of flux measurements by up to 30\%. A method has been developed to correct detector transients during data reduction \cite{fadda16,sutter22} for the R1 band. We adopted this method and reprocessed the data to correct the detector transients in the B2B band.  

\noindent\textbf{PSF-Matching}

\begin{figure}[htbp]
\centering
\includegraphics[width=11cm]{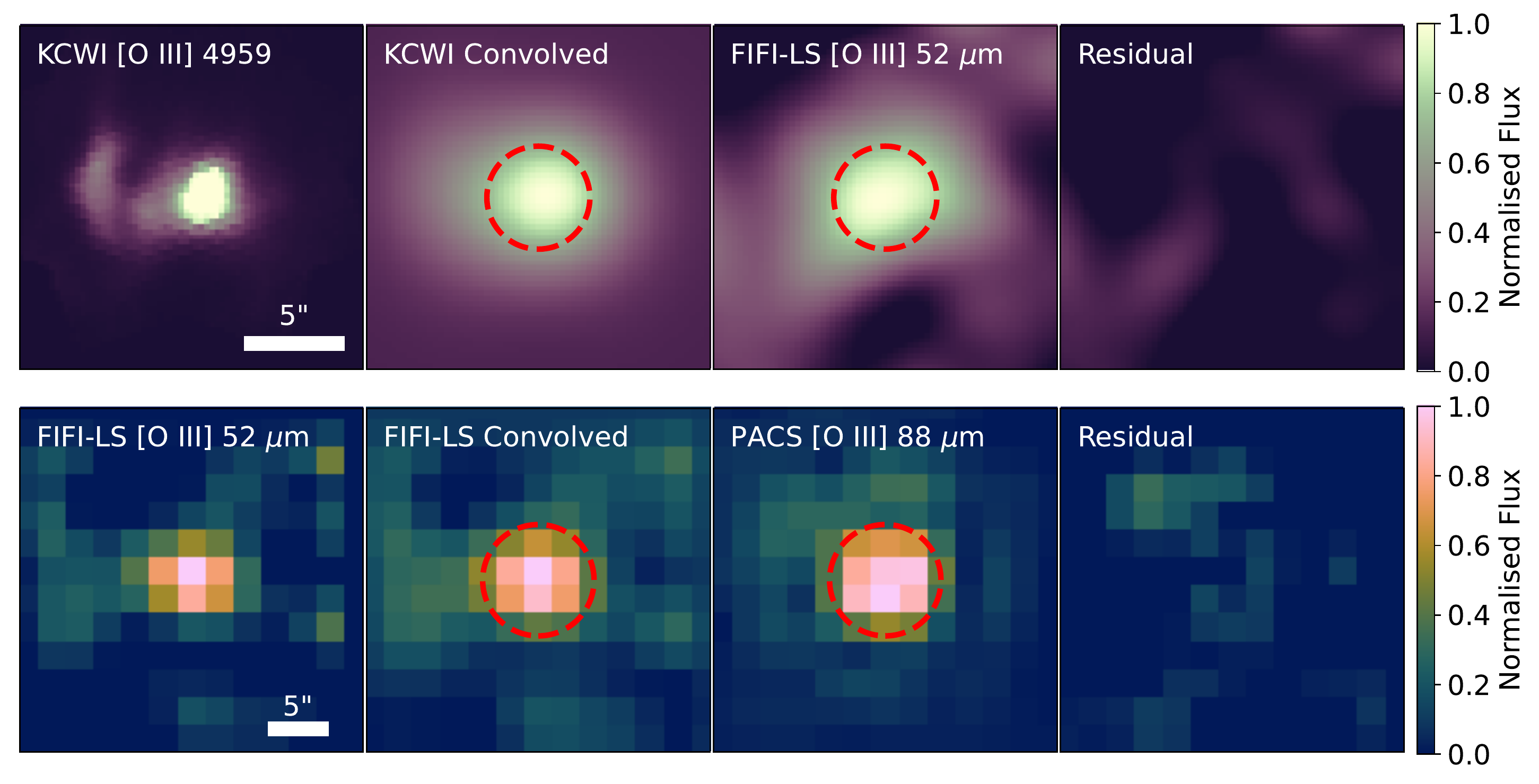}
\caption{  PSF matching between KCWI and FIFI-LS (top), and between FIFI-LS and PACS (bottom). From left to right: the pseudo-narrowband (PNB) emission line images with smaller PSF before convolving the Gaussian kernel; the same PNB images after convolving with the Gaussian kernel; the PNB images with larger PSF to be matched; and the residual between the convolved images and the images with larger PSF. The red dashed circles indicate the apertures from which the spectra are extracted.}
\label{fig:psf_matching}
\end{figure}

In order to measure the flux ratios consistently from all instruments, we convolved all data to the same spatial resolution, using 2D Gaussian kernels to match lower-resolution data. 
Due to the PACS spatial resolution being comparable to the KCWI FoV, we obtain more robust results by conducting the PSF matching separately: 1) between KCWI and FIFI-LS, and 2) between FIFI-LS and PACS (Fig. \ref{fig:psf_matching}). We extracted the continuum-subtracted pseudo-narrowband (PNB) images near \Oiii{4959}, \Oiii{52~\um}, and \Oiii{88~\um}. The spectral windows covering the line emission were determined by eye to include the wavelength elements with flux elevated above the adjancent continuum level. The median flux in these adjacent pixels was then subtracted from the line image to remove the continuum. 
We convolved the images with smaller PSF (higher resolution) with an axisymmetric kernel of varying FWHM in order to fit to the lower-resolution data. 
The best-fit FWHMs of the convolution kernels are 5\secpoint9 between KCWI and FIFI-LS and 4\secpoint9 between FIFI-LS and PACS. These are in good agreement with expectations based on the seeing (with KCWI) and diffraction limits of FIFI-LS and PACS. Finally, we convolved the full data cubes with the best-fit kernels derived from the PNB images to produce PSF-matched data. 

\begin{figure}[htbp]
\centering
\includegraphics[width=8cm]{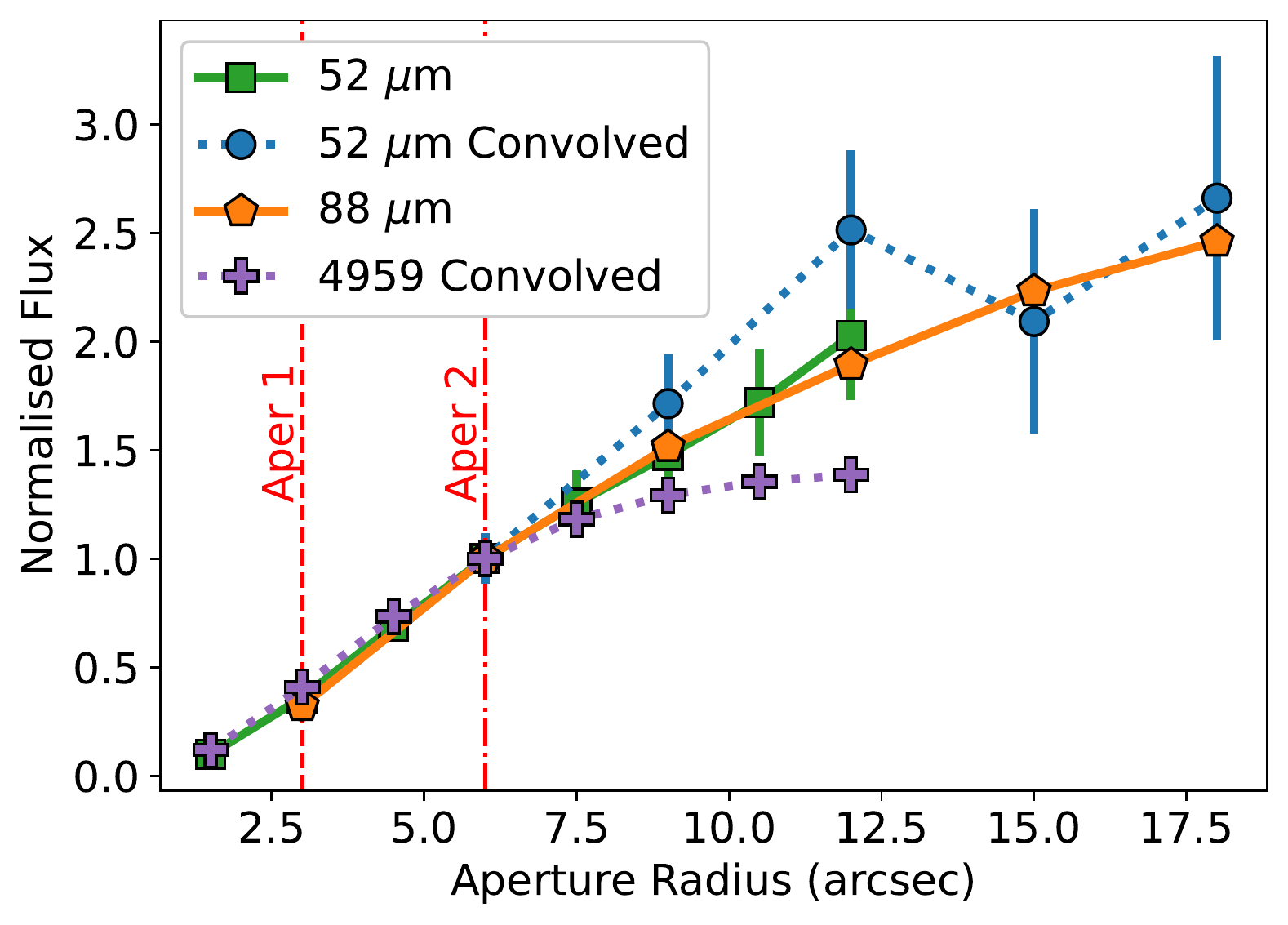}
\caption{ Spatial curves of growth (CoG) of the \Oiii\ emission lines after PSF matching. 
All CoGs are consistent within radii $\lesssim$7\arcsecond. Larger radii correspond to regions beyond the KCWI field of view, causing the plateau in the \Oiii{4959} CoG from KCWI. 
The far-IR CoGs remain consistent with each other at larger radii corresponding to their larger fields of view. We use Aperture 1 ($r = 3''$, dashed vertical line) to extract spectra from the convolved KCWI and FIFI-LS data, and Aperture 2 ($r = 6''$, dash-dotted vertical line) for the convolved FIFI-LS and PACS data. }
\label{fig:cog}
\end{figure}

We verify the reliability of our PSF matching by measuring the spatial curve of growth (CoG) -- the encircled flux as a function of aperture radius (\Rap) -- of the \Oiii\ emission lines. The CoG of all \Oiii\ emission lines presented in Fig. \ref{fig:cog} are consistent with each other at $\Rap \lesssim 7''$. The KCWI \Oiii~4959 CoG flattens at $\Rap \gtrsim 8''$, as larger apertures are beyond the KCWI FoV. The far-IR CoGs continue to increase in flux at larger $\Rap$ and show good agreement, although the convolved FIFI-LS \Oiii~52~\um\ has lower signal-to-noise ratio at $\Rap \gtrsim 10''$ (due to increased noise near the edge of the FIFI-LS FoV). 
We note that the eastern edge of the PACS FoV includes a fainter but previously known star-forming region, NGC~2366-II \cite{micheva17}. This highlights the necessity to conduct PSF matching and motivates us to choose a relatively small aperture radius to ensure that the flux is dominated by the central region observed by all three instruments. Furthermore, the S/N of the spectra is also maximised with a moderately small aperture radius. We therefore used $\Rap = 3''$ for the matched cubes between KCWI and FIFI-LS and $\Rap = 6''$ for the matched cubes between FIFI-LS and PACS. The increase in \Rap\ in the latter scenario is motivated by the larger pixel and PSF sizes, while maintaining a near-optimal S/N. For purposes of comparing flux ratios from KCWI and PACS, we scale the PACS \Oiii~88~\um\ flux based on the FIFI-LS \Oiii~52~\um\ flux ratios in the two scenarios, although our results do not strictly require this comparison.

\noindent\textbf{Line Flux Measurements and Uncertainties}

Most line fluxes were measured by fitting the Voigt profile to the 1D spectra extracted from the PSF-matched data cubes. Before fitting, the continuum flux was removed using a linear function that fits the blue and red sides of the continuum spectra in the following range: [-1000, -400] \kms\ and [400, 1000] \kms\ for KCWI, [51.65, 51.78] \um\ and [51.90, 52.01] \um\ for FIFI-LS, and [88.01, 88.20] \um\ and [88.57, 88.81] \um\ for PACS. 
For all emission lines, the total residual fluxes after subtracting the Voigt profile are $< 1\%$ of the line fluxes. Broad emission wings associated with nebular emission lines have previously been detected in Mrk~71 \cite{komarova21}. However, their velocities extend $> 1000$~\kms, and their peak flux is $< 1\%$ of the primary component. These features are potentially detected in the residual spectra of KCWI and PACS, but because of their high velocities, they do not contribute significantly to the residual fluxes. 

Fluxes of the \Oiirec~4649 recombination multiplet and the \Oii~3726, 3729 doublet were measured by fitting multicomponent Gaussian models. For \Oiirec\ recombination lines, the line ratios among the lines in the \Oiirec~4649 multiplet are sensitive to $n_e$. Instead of fitting the lines independently, we set $n_e$ as a free parameter that determines the line ratios. The measured fluxes are provided in Table \ref{tab:fluxes}.

Flux uncertainties for our analysis include both random and systematic errors. The random error, dominated by photon-counting noise for the KCWI data and by thermal noise for the FIFI-LS and PACS data, is estimated from the error spectra which are extracted alongside the 1D flux spectra from the reduced data cubes. However, because of the PSF convolution and oversampling, the pixel correlation is not negligible. We estimate the spatial  {and spectral} covariance between adjacent pixels in the following way \cite{law16}. The error spectra are first extracted assuming no pixel correlation. These error spectra were then scaled to match  {the standard deviation of the fluxes randomly drawn from a spectral width equal to the FWHM of the line profile} in regions where no emission lines were present, and the flux errors of the emission lines were estimated based on the scaled error spectra. 

The systematic uncertainty is dominated by flux calibration. For KCWI, we compared our observations with the archival F438W imaging (PI: B. James \cite{james16}) with the Wide Field Camera 3 onboard the Hubble Space Telescope (HST/WFC3). The F438W filter covers approximately 3895--4710 \AA, which includes many of the key emission lines used in this study. We constructed a pseudo-F438W image from the KCWI data, and smoothed the WFC3 image to match the KCWI seeing following the PSF-matching approach outlined above. The resulting KCWI flux is $\simeq 7\%$ higher than the HST flux. This difference is likely due to the variation of atmospheric transparency over the course of the night, and was removed by rescaling the KCWI flux. The RMS of the difference between the flux-calibrated KCWI spectra of the photometric standards and the DRP model spectra is $\sim 3\%$. We adopt this flux-calibration uncertainty as the dominant systematic for our KCWI data, and it is included in the reported metallicity uncertainties.

For the far-IR data, after the PACS detector transient correction, the flux calibration uncertainty between FIFI-LS and PACS has been reported as $\sim 15\%$ in the PACS R1 band \cite{sutter22}. This uncertainty is based on overlapping observations from both instruments on the [\ion{C}{ii}]~158~\um\ line. We confirmed this uncertainty on the [\ion{C}{ii}] emission of a handful of star-forming regions, including Mrk~71, using our PSF-matching routine. We suspect that this uncertainty is reduced in the PACS B2B band because the detector transient effect is much less prominent. Indeed, we analysed a nearby star-forming object with overlapping \Oiii~88~\um\ observations from FIFI-LS and the PACS B2B band and found that the flux difference, after PSF-matching, is $\lesssim 10\%$. 
For this study we conservatively adopt a flux calibration uncertainty of 15\% between the FIFI-LS and PACS observations, and we quadratically divide this such that the absolute flux uncertainty of FIFI-LS and PACS is $\sim 10\%$ each. This uncertainty was then added quadratically to the thermal noise discussed above. 

\noindent\textbf{Reddening Correction}

The internal dust reddening of Mrk~71 was measured by comparing the observed H Balmer fluxes with the intrinsic flux ratios for Case B recombination, assuming $n_e = 100~\mathrm{cm}^{-3}$ and $T_e = 10,000~\mathrm{K}$\footnote{The Balmer line ratios are relatively insensitive to $T_e$ and $n_e$. For $T_e = 15,000~\mathrm{K}$ and $n_e = 200~\mathrm{cm}^{-3}$, the H$\beta$/H$\alpha$ ratio only changes by $<1\%$.}. The H Balmer lines included here range from H$\beta$ to H$\kappa$, except for H$\epsilon$ and H$\zeta$ which are blended with other features. 
We assumed a Milky Way extinction curve \cite{cardelli89} with $R_V = 3.1$. 
We additionally consider the contribution from underlying stellar absorption, which may be non-negligible especially for the higher-order Balmer lines. Indeed, we observe that higher-order Balmer lines (at shorter wavelengths) imply larger correction factors which cannot be explained by increasing $R_V$, and instead suggest non-negligible stellar absorption. 
 {When fitting the measured Balmer line fluxes, we included the stellar absorption component by assuming that the absorption equivalent widths (\wabs) are constant for all Balmer lines. The best-fit value is $\wabs = 0.6 \pm 0.7$~\AA. Although this result suggests that the absorption is not significant ($\sim 1\sigma$), the wings of stellar absorption features are apparent in visual inspection of the deep higher-resolution KCWI spectra, from which we estimate $\wabs \sim 0.5$~\AA\ from H$\delta$ and H$\eta$. We conservatively adopt the best-fit uncertainty for our analysis. }

We also analysed the fluxes previously reported by \cite{esteban09} from high-resolution slit spectroscopy with Keck/HIRES, with the same fitting process. Due to the inclusion of the H$\alpha$ flux, the detection of \wabs\ is $>6\sigma$\footnote{The absolute value of \wabs\ is not reliable, since the continuum flux was not available in E09, and is instead inferred from our KCWI data.} in the HIRES data, while the total extinction correction \chb is consistent with our KCWI measurement to well within $1\sigma$. The presence of \wabs\ is however in contradiction with the previously reported $\wabs = 0$ \cite{esteban09}. We suspect that the difference is caused by the use of only bright lines in earlier work (H$\alpha$--H$\epsilon$), which are less sensitive to \wabs. 
The best-fit \chb\ and \wabs\ are included in Table \ref{tab:properties}. The uncertainty in \chb\ is included in our reported metallicity measurements. 

\begin{figure}[htbp]
\centering
\includegraphics[width=5cm]{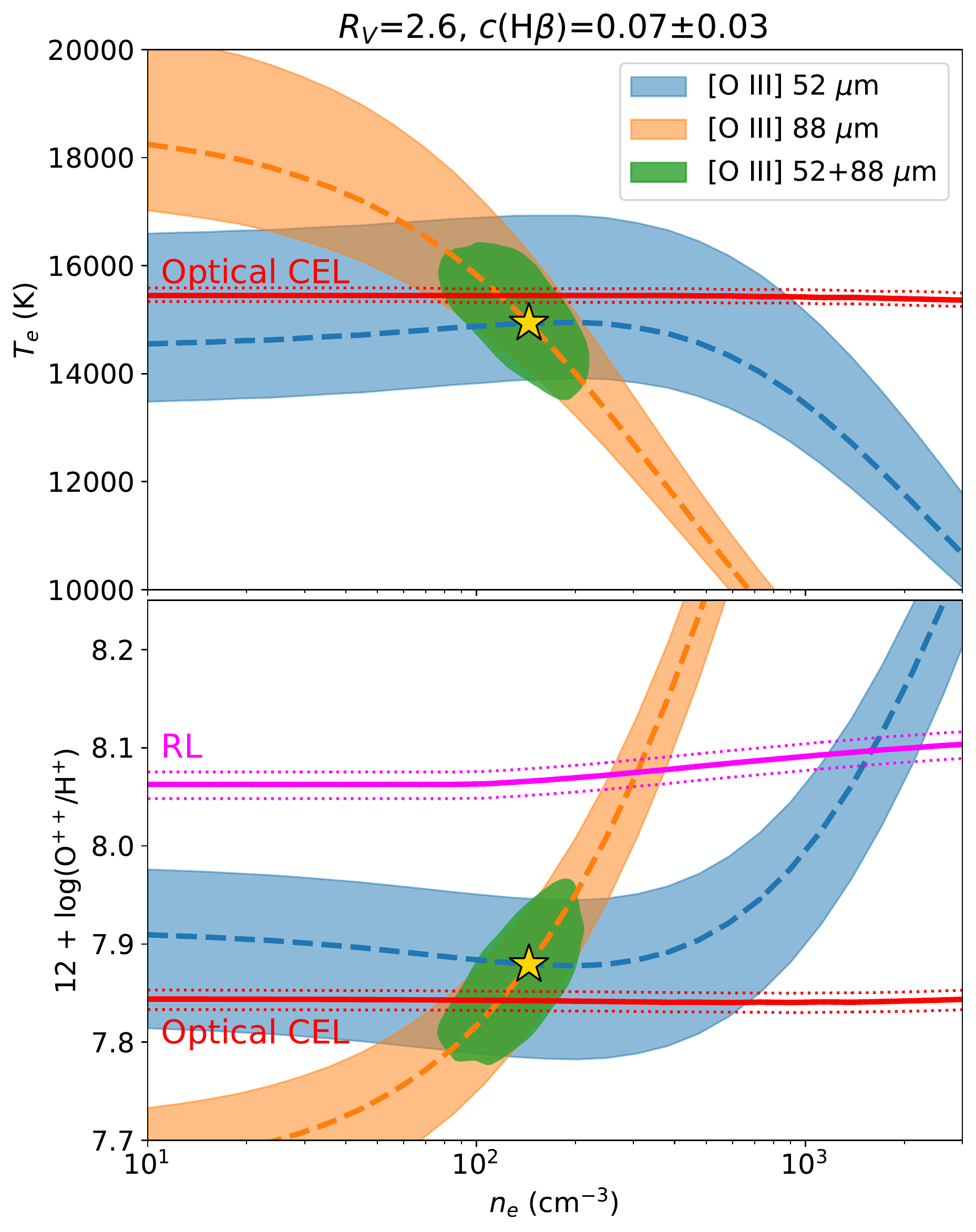}
\includegraphics[width=5cm]{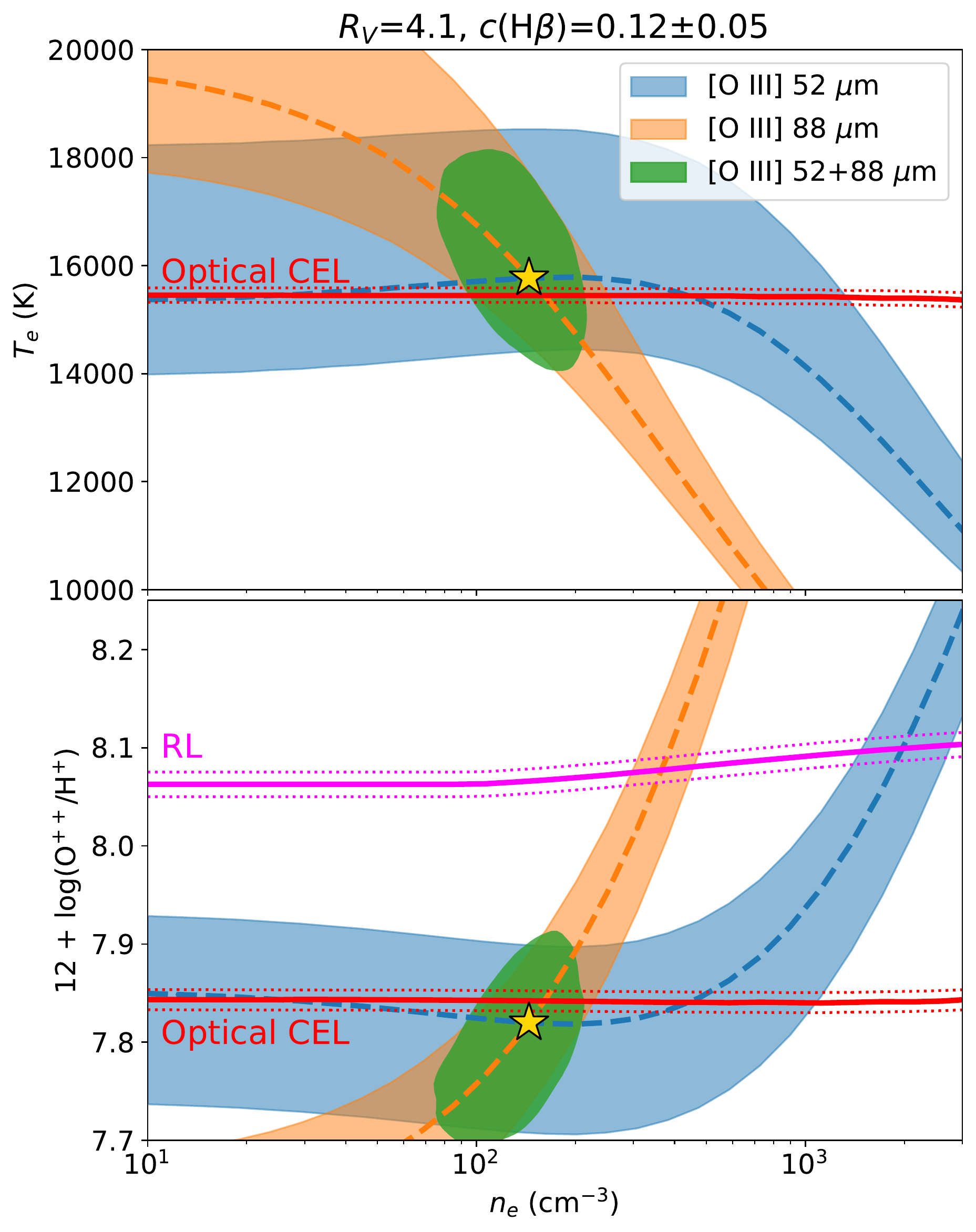}
\includegraphics[width=5cm]{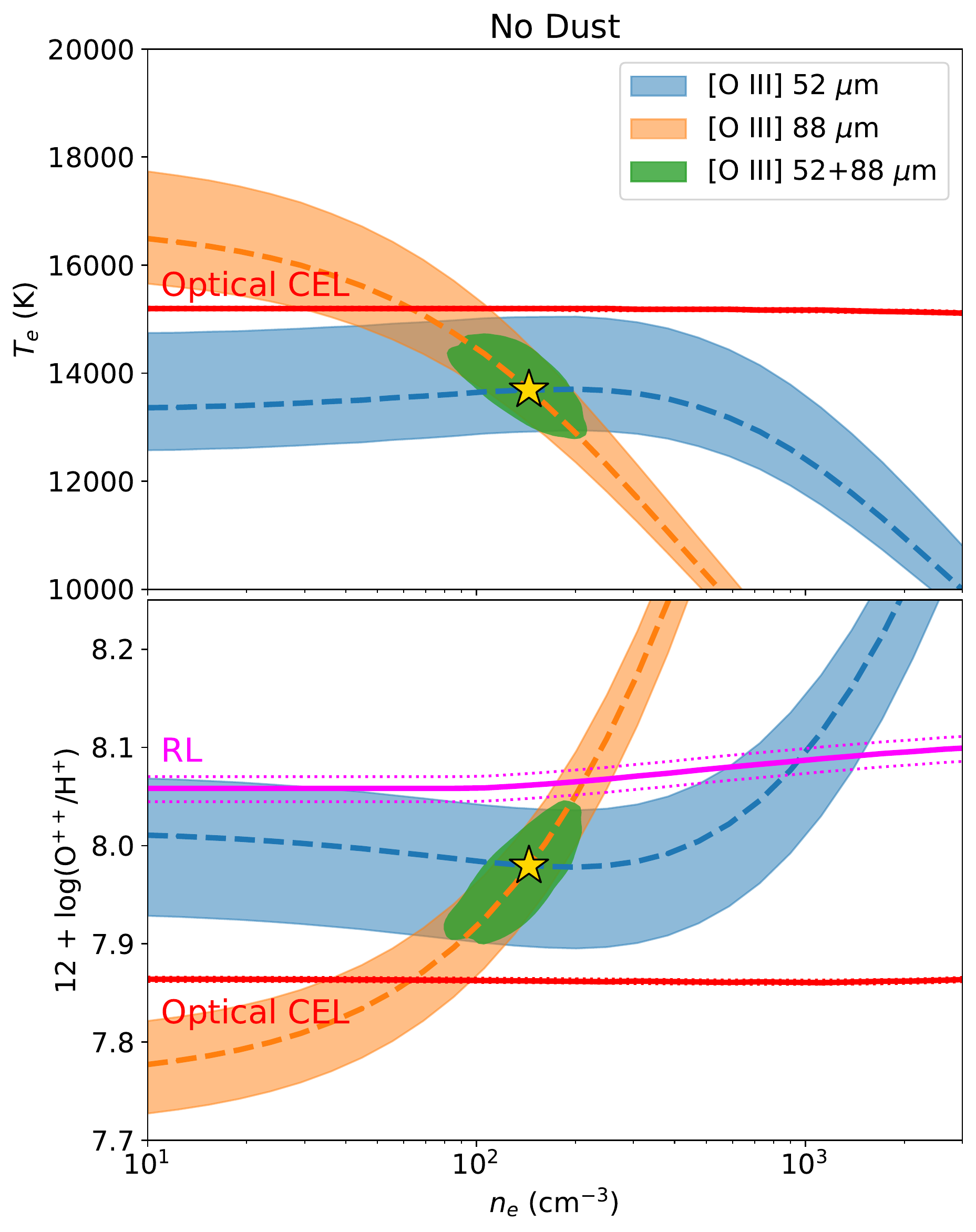}
\caption{   The impact of alternative extinction curves ($R_V = 2.6$, $R_V = 4.1$, and no dust) on the metallicity measurements. Different $R_V$ values slightly offset the far-IR metallicities, but the inconsistency between the far-IR and RL metallicities does not change significantly. The RL metallicity is still different from the far-IR metallicity by $ > 1\sigma$ even if no extinction correction is applied.  }
\label{fig:alt_dust}
\end{figure}

To further investigate potential systematic uncertainties in the reddening correction, we explored alternative $R_V$ values. We repeat the extinction correction processes mentioned above and remeasure the metallicities with $R_V$ values of 2.6 and 4.1 (see Fig. \ref{fig:alt_dust}  {and Table \ref{tab:alt_dust}}).  {An even higher $R_V$ is possible \cite{urbaneja2017}, but only increases the correction factors for optical fluxes and enhances the inconsistency between the far-IR and RL metallicities.} A low $R_V$ reduces the significance of the discrepancy. However, $R_V \lesssim3$ has rarely been reported in the literature for such galaxies. In Fig. \ref{fig:alt_dust}, we also include an unlikely scenario of no dust extinction (i.e., \chb~$=0$).  {Without extinction correction, the joint metallicity constraint of the far-IR \Oiii\ to H$\beta$ ratios is closer to the RL metallicity compared to the optical CEL metallicity--i.e. the temperature fluctuation explanation is favoured. However, the discrepancy between far-IR metallicity and RL metallicity is still $> 1\sigma$, such that the implied temperature fluctuations are still not large enough to account for the ADF even under this extreme assumption. }

\noindent\textbf{Temperature and Metallicity}

The electron temperature ($T_e$) and metallicity of Mrk~71 were measured from the reddening-corrected flux ratios using \textsc{PyNEB} \cite{luridiana15}  {using the default collisional strengths \cite{storey14} for \Oiii\ CEL and recombination coefficients \cite{storey17}} for \Oiirec\ RL. The uncertainties were derived from a Markov Chain Monte-Carlo (MCMC) approach accounting for the random (photon-counting or thermal) noise estimated from the data reduction pipelines, systematic flux calibration uncertainties, and uncertainty in the reddening correction. 

For optical CEL, the critical densities are $n_\mathrm{crit} > 10^3~\mathrm{cm}^{-3}$, such that their emissivities are nearly independent of $n_e$ in the range of our interest. Therefore, $T_e$ (CEL) can be measured from the \Oiii~4363 / 4959 ratio with negligible density dependence. The O$^{++}$/H$^{+}$ metallicity was calculated from the \Oiii~4959 / H$\beta$ flux ratio using the emissivities determined by $T_e$ (CEL). In contrast the \Oiirec~4649 RL multiplet lines (including \Oiirec~4638, 4641, 4649, 4650, and 4661)  {are dominated by recombination, with little to no contributions from fluorescent emission \cite{esteban04}}, and are only weakly sensitive to $T_e$. Therefore, we used $T_e$ (CEL) to calculate the RL emissivity and used the RL / H$\beta$ ratio to calculate the RL metallicity. The relative flux ratios of these emission lines are sensitive to $n_e$ \cite{storey17}, and we performed an MCMC fit to the RL spectra (Fig. \ref{fig:all_spec}) to derived $n_e$ (RL) and its uncertainty. 

Unlike optical CEL, the far-IR \Oiii\ lines have relatively small critical densities, such that their emissivities depend on $n_e$ (and weakly on $T_e$). A key aspect of this work is the combined measurements of \Oiii~52~\um\ and 88~\um, which have different critical density, provide a direct measurement of $n_e$ appropriate for the far-IR lines. We thus obtain good joint constraints on $n_e$, $T_e$, and metallicity. 
We first performed an MCMC analysis on the reddening-corrected and aperture-matched \Oiii~4959, 52~\um, 88~\um\ fluxes, to derive the posterior likelihood in the $T_e$ and $n_e$ space. This posterior is then used as the prior to fit the (\Oiii~52~\um + 88~\um) / H$\beta$ ratio in a second MCMC, to obtain the far-IR O$^{++}$/H$^+$ metallicity (Fig. \ref{fig:oh_metallicity}). 

 {We also provide the total O/H metallicity in Table \ref{tab:properties}, although we note that the ADF measurements are based strictly on the $\mathrm{O}^{++}$ ion. The O/H metallicity was estimated as $(\mathrm{O}^{+} + \mathrm{O}^{++})/\mathrm{H}^{+}$, assuming that O$^{+}$, O$^{++}$, and H$^+$ make up the entire populations of O and H. We calculated O$^{+}$/H$^+$ metallicity from the extinction-corrected \Oii~3726, 3727 fluxes, by assuming $T_e(\Oii) = 0.71 \times T_e(\Oiii~\mathrm{CEL}) + 3,050~\mathrm{K}$ \cite{esteban09}.   }

\begin{table}[h]
\begin{center}
\begin{minipage}{174pt}
\caption{Emission line fluxes of Mrk~71}\label{tab:fluxes}%
\begin{tabular}{@{}lc@{}}
\toprule
Emission Line & Observed Flux\footnotemark[1]  \\
& ($10^{-14}$~erg~s$^{-1}$~cm$^{-2}$) \\
\midrule
\multicolumn{2}{l}{\textit{Instrument: KCWI}} \\
\Oiii~4363 & $6.24 \pm 0.03$\\
\Oiii~4959 & $104.36 \pm 0.07$ \\
\Oiirec~4649 Multiplet\footnotemark[2] & $0.061 \pm 0.002$\\
\Oii~3726 & $9.51 \pm 0.07$\\
\Oii~3729 & $12.32 \pm 0.07$ \\
H$\beta$ & $44.05 \pm 0.05$\\
H$\gamma$ & $20.34 \pm 0.04$\\
H$\delta$ & $10.47 \pm 0.03$ \\
H$\eta$ & $2.94 \pm 0.03$ \\
H$\theta$ & $2.17 \pm 0.03$ \\
H$\iota$ & $1.54 \pm 0.04$ \\
H$\kappa$ & $1.11 \pm 0.04$\\
\midrule
\multicolumn{2}{l}{\textit{Instrument: FIFI-LS}} \\
\Oiii~52~\um & $38 \pm 4$\\
\midrule
\multicolumn{2}{l}{\textit{Instrument: PACS}\footnotemark[3]} \\
\Oiii~88~\um & $42.3 \pm 0.4$ \\
\botrule
\end{tabular}
\footnotetext[1]{The uncertainties quoted here only include random noise. See Methods for flux calibration uncertainties. }
\footnotetext[2]{Total flux includes: \Oiirec~4638, 4641, 4649, 4650, and 4661. Measured from the deep BM exposures. }
\footnotetext[3]{The flux extracted from a 6\arcsecond\ aperture has been scaled by a multiplicative factor of 0.447 to match the 3\arcsecond\ aperture used for KCWI (see Methods).}
\end{minipage}
\end{center}
\end{table}

\begin{table}[h]
\begin{center}
\begin{minipage}{500pt}
\caption{Properties of Mrk~71 from Alternative Extinction Models$^1$}\label{tab:alt_dust}%
\begin{tabular}{@{}lccc@{}}
\toprule
Physical Properties & $R_V = 2.6$ & $R_V = 4.1$ & No Dust \\
\midrule
$\chb$ & $0.07 \pm 0.03$ & $0.12 \pm 0.05$ &  \\
\wabs & $0.6 \pm 0.7~\mathrm{\AA}$ & $0.6 \pm 0.7~\mathrm{\AA}$ & \\
$T_e$ (far-IR) & $14,900_{-900}^{+1,700}~\mathrm{K}$ & $15,800_{-1,200}^{+2,400}$ & $13,700_{-700}^{+1,200}$ \\
$T_e$ (CEL) &   $15,400 \pm 100~\mathrm{K}$ & $15,400 \pm 100~\mathrm{K}$ & $15,200 \pm 30~\mathrm{K}$  \\
$12 + \log(\mathrm{O}^{++} / \mathrm{H}^+)$ (far-IR) & $7.88_{-0.10}^{+0.06}$ & $7.82_{-0.11}^{+0.07}$ & $7.98_{-0.08}^{+0.05}$\\
$12 + \log(\mathrm{O}^{++} / \mathrm{H}^+)$ (CEL)\footnotemark[2] & $7.842_{-0.010}^{+0.009}$ & $7.842 \pm 0.010$ & $7.862 \pm 0.002$ \\
$12 + \log(\mathrm{O}^{++} / \mathrm{H}^+)$ (RL)\footnotemark[2] & $8.07 \pm 0.01$ & $8.07 \pm 0.01$ & $8.06 \pm 0.01$\\
$12 + \log(\mathrm{O} / \mathrm{H})$ (far-IR)\footnotemark[3] & $7.91_{-0.09}^{+0.06}$ & $7.85_{-0.10}^{+0.07}$ & $8.00_{-0.08}^{+0.05}$ \\
$t^2$ (between far-IR and CEL) & $0.018 \pm 0.038$ & $0.005 \pm 0.042$ & $0.06 \pm 0.04$  \\
$t^2$ (between CEL and RL) & $0.096 \pm 0.008$ & $0.096 \pm 0.008$ & $0.084 \pm 0.006$  \\
\botrule
\end{tabular}
\footnotetext[1]{Properties that do not vary with the extinction law assumptions are omitted from this table.}
\end{minipage}
\end{center}
\end{table}

\begin{figure}[htbp]
\centering
\includegraphics[width=8cm]{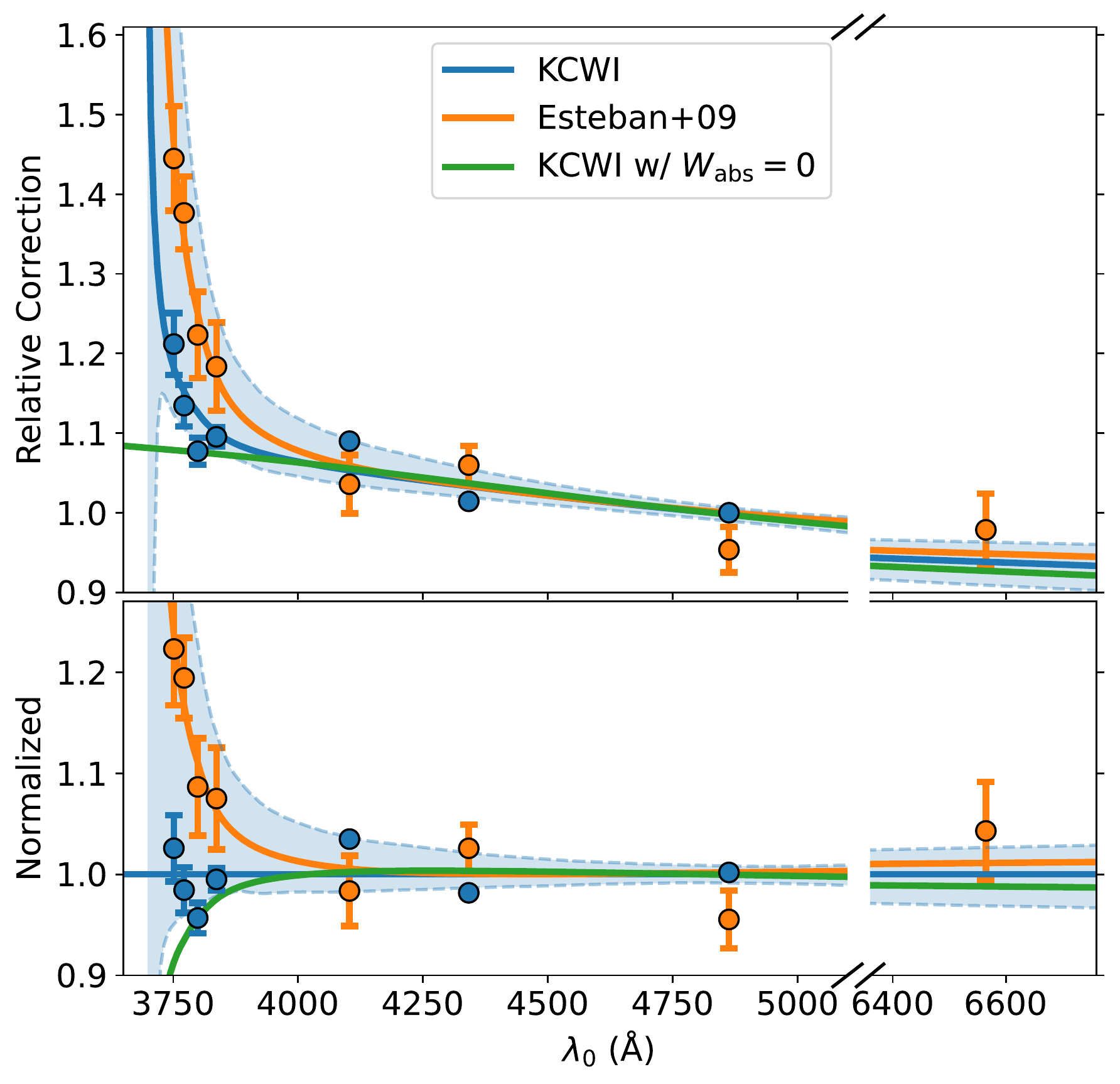}
\caption{  Best-fit reddening correction curves. Top: The required correction factor is measured as the intrinsic Balmer flux ratios (normalised to H$\beta$) compared to the observed ratios. Bottom: Flux ratios after extinction correction, where the best-fit model based on the KCWI model (blue) was used. Different coloured points indicate the origin of the fluxes, with blue indicating our own KCWI dataset and orange indicating the E09 \cite{esteban09} dataset. The green curve is the best-fit extinction model for the KCWI data without including contributions from stellar absorption. }
\label{fig:dust_correction}
\end{figure}

\bibliography{sn-bibliography}

\end{document}